\documentclass{jaa}
\usepackage{natbib}

\usepackage[section]{placeins}
\usepackage{array,longtable}
\usepackage{amsmath}
\usepackage{graphicx}

\usepackage{xcolor}
\usepackage{gensymb}
\usepackage[percent]{overpic}
\usepackage[colorlinks=true,
            linkcolor=blue,
            citecolor=blue,
            urlcolor=blue]{hyperref}
\usepackage{gensymb}
\begin{document}\sloppy

\title{First uGMRT Constraints on Rotation Measure in the Radio-Quiet Seyfert Galaxy NGC~4235}

\author{Salmoli Ghosh\textsuperscript{1}, Preeti Kharb\textsuperscript{1}, and Janhavi Baghel\textsuperscript{2, 3}}
\affilOne{\textsuperscript{1}National Centre for Radio Astrophysics - Tata Institute of Fundamental Research (NCRA-TIFR), S. P. Pune University Campus, Ganeshkhind, Pune 411007, India\\}
\affilTwo{\textsuperscript{2}Department of Physics, University of Pretoria, Hatfield, Pretoria, 0028, South Africa\\}
\affilThree{\textsuperscript{3}Inter-University Institute for Data Intensive Astronomy, and Department of Physics, University of Pretoria, Pretoria, South Africa

}

\twocolumn[{
\maketitle
\corres{salmoli@ncra.tifr.res.in}
\msinfo{}{}

\begin{abstract}
We present a multi-frequency polarimetric study of the radio-quiet Seyfert galaxy NGC 4235 using the upgraded Giant Metrewave Radio Telescope (uGMRT), aimed at constraining the Faraday rotation properties of its kpc-scale radio emission. This marks the first-ever analysis of polarization across two frequency bands of the uGMRT {and the first constraints
on rotation measure} (RM) at lower frequencies for a weak radio source ($17.1\pm1.7$ mJy at 1.3 GHz, $27.7\pm2.8$ mJy at 663 MHz). We detect linearly polarized radio emission associated with the `core' and the extended lobes of NGC~4235. Broadband polarimetric analysis yields an RM value of {$\sim200$ rad~m$^{-2}$} in the `core', indicating Faraday rotation by magnetized plasma external to the synchrotron-emitting regions. Using these RM constraints, we estimate an average electron density of the Faraday-rotating medium of {$\sim3\times10^{-2}$~cm$^{-3}$} for a characteristic perturbation scale of {$\sim0.3$ parsec within an assumed magnetized region of size $\sim1.3$ kpc}, consistent with a diffuse magnetized sheath or AGN-driven wind. These results extend classical Faraday rotation diagnostics to the low-power, radio-quiet AGN regime and demonstrate the potential of uGMRT broadband polarimetry for probing magnetized environments in and around such AGN.
\end{abstract}

}]

\section{Introduction}

Active galactic nuclei (AGN) are accreting supermassive black holes (SMBHs), driving powerful electromagnetic emission. While $\sim$20\% of the AGN launch strong, extended relativistic jets, called as the radio-loud (RL) AGN, the remaining $80\%$ exhibit weak radio emission mostly confined within the host galaxy and are called radio-quiet (RQ) AGN \citep[radio loudness parameter $R$ is defined as the ratio of flux density at 5 GHz to that at optical B-band; $R\leq10$ for RQ AGN;][]{Kellermann1989}. The origin of radio emission in RQ AGN is debated, with major contenders being the AGN-driven jet or wind \citep{Baum1993, Kukula1996, Kharb2026}, starburst-driven wind \citep{Condon1992, Thean2000}, coronal activity \citep{Laor2008, Panessa2013}, or outflow-driven shocks \citep{Crenshaw2000,Zakamska2014}, summarised in the review paper of \cite{Panessa2019}. Interestingly, polarization studies in the radio regime provide a direct probe of the magnetic (B-) field geometry, the physical conditions of the emitting plasma, and the magneto-ionic medium along the line of sight. This can offer vital clues regarding the nature and origin of outflows and AGN `feedback' from RQ AGN \citep{Silpa2021a,Kharb2023, Ghosh2025, Ghosh2025IAU}. 

Radio emission in AGN, predominantly arising from the highly linearly polarized synchrotron emission mechanism, gets partly or fully depolarized as it reaches us \citep[see][]{Burn1966,Shneider2014}. As linearly polarized waves propagate through magnetized plasma, the plane of polarization rotates due to the Faraday effect. {If different parts of the emitting region or regions within the telescope beam or within the observing frequency range produce different amounts of rotation, the polarization vectors partially cancel, due to internal, beam, and bandwidth depolarization, respectively \citep[][Chapter 3]{Contopoulos2015}. The final observed polarization angle ($\phi_{obs}$) provides diagnostics of the rotation due to the presence of the magneto-ionic medium.} The electric vector position angle (EVPA) changes from the intrinsic angle of polarization ($\phi_{int}$) by rotation measure (RM) times wavelength squared ($\lambda^2$) as:
\begin{equation}
    \phi_{obs}=\phi_{int}+\mathrm{RM} \lambda^2
    \label{eq:chiRMlambda}
\end{equation}
{RM depends on the electron density ($n_e$) and the magnetic field ($B_\parallel$) integrated along the line of sight as:}
\begin{equation}
\mathrm{RM}
= \frac{e^{3}}{2\pi m_{e}^{2} c^{4}}
\int_{0}^{L} n_{e} (l)\, B_{\parallel} (l)\, dl
\label{eq:RMintegral}
\end{equation}
{ By assuming that the magneto-ionic medium has a uniform electron density and magnetic field along a path length $L$, Equation \ref{eq:RMintegral} becomes}
\begin{equation}
    \mathrm{RM}~(\mathrm{rad\, m^{-2}})=812\,n_e (\mathrm{cm^{-3}})\,B_\parallel\ (\mathrm{\mu G}) L\,(\mathrm{kpc})
    \label{eq:RM}
\end{equation} 
Measurements of RM therefore offer a powerful diagnostic of both the intrinsic magneto-ionic environment near the AGN and the intervening media.

At low radio frequencies ($<1$ GHz), Faraday rotation effects start to dominate, often resulting in depolarization and large rotations in angles, limiting the number of {robustly detected
polarized} emissions at MHz frequencies \citep{Lenc2017}. Moreover, the polarization signals tend to be much weaker in RQ AGN compared to the radio-loud ones. Such low polarization makes it difficult to distinguish intrinsic AGN polarization from foreground Galactic contributions or instrumental systematics. The upgraded Giant Metrewave Radio Telescope (uGMRT) has been found to be able to detect polarization at $>1.5\%$ level at 550 $-$ 850~MHz \citep{Kharb2023arXiv}.

NGC~4235 is a nearby, nearly edge-on spiral galaxy ($z\approx0.0075$) hosting a Seyfert 1.5 nucleus \citep{Gallimore2006}, with a black hole mass of $2.5\times10^7$ M$_\odot$ \citep{Dong2006}. It is a low-luminosity RQ AGN with a radio flux density of $\sim11$ mJy at 5 GHz \citep{Gallimore2006}, $\sim15.5$ mJy at 1.4 GHz \citep{Condon1998}, and $\sim13$ mJy at 660 MHz \citep{Kharb2016}. Very Long Baseline Array (VLBA) imaging revealed a compact nuclear radio source. Low-frequency observations with the GMRT at 325 and 610 MHz revealed kpc-scale double-lobed radio structure extending to $\sim24$ kpc, consistent with non-thermal AGN-powered emission rather than star-formation processes \citep{Kharb2016}. The spectral index ($\alpha$, with flux density F being proportional to frequency $\nu$ as F $\propto\nu^\alpha$) imaging between 325 and 610 MHz reveals spectral indices of $\approx -0.3$ to $-0.6$ in the lobes, suggesting that the lobes are not old or `relic'. We note that the `relic' emission detected beyond the `active' lobes in NGC~4235 by \citet{Kharb2016} is not picked up in our current images, likely due to relatively less sensitivity (r.m.s. noise of $\sim40~\mu$Jy~beam$^{-1}$ in the previous image versus $\sim60~\mu$Jy~beam$^{-1}$ in the new image). The alignment of the pc-scale jet and the kpc-scale radio lobe reinforces an AGN jet origin for the extended radio structures in NGC~4235.

In this paper, we present a polarization study of the source at low radio frequencies (663 MHz and 1.3 GHz) with the uGMRT. {Combined with} the 10~GHz data from the Karl G. Jansky Very Large Array \citep[VLA, Project 19B-198;][]{Ghosh2025}, we have attempted to constrain the RM for the first time at the centre of this galaxy. This study also provides the first wide-band RM estimate in an RQ AGN using multi-instrument polarimetric observations, exploring the low-frequency regime with the uGMRT.

\section{Observations \& Data Reduction}\label{sec:obs}
We have observed the source NGC 4235 in full polar mode, projects 45\_086 and 46\_086, at 1.3~GHz and 663~MHz, respectively. The uGMRT at $\sim$660 MHz (Band-4) has native linear feeds along with a `quadrature hybrid,' aka polarizer, converting linear to circularly polarized signals. At $\sim$1.3~GHz (Band-5), due to the absence of such a polarizer, linearly polarized signals are received.
Solving the polarization equations in the circular basis is more straightforward than in the linear basis \citep[see Section 3 in][]{Hales2017}. In the circular basis, the Stokes $Q$ and $U$ parameters are derived solely from the cross-hand correlations of $RL$ and $LR$, and instrumental gain errors can be decoupled from the polarization signal, making calibration of astrophysical signals (especially with zero circular polarization) easier. 

We use our own (publicly available) uGMRT-based polarization data reduction pipelines{\footnote{\href{https://github.com/astrouser-salm/radio-imaging/blob/main}{https://github.com/astrouser-salm/radio-imaging}}} using Common Astronomical Software Application ({\tt CASA}), for reducing uGMRT Band-4 (circular basis) and Band-5 (linear basis) data. We note that the validation of the uGMRT Band-5 polarimetric calibration strategy across a wider range of observing conditions is ongoing, and further applications will help establish its robustness.

\begin{figure*}
  \centering
  \vspace{0.5cm}
  \begin{overpic}[width=11cm, trim= 0 80 40 99,clip]{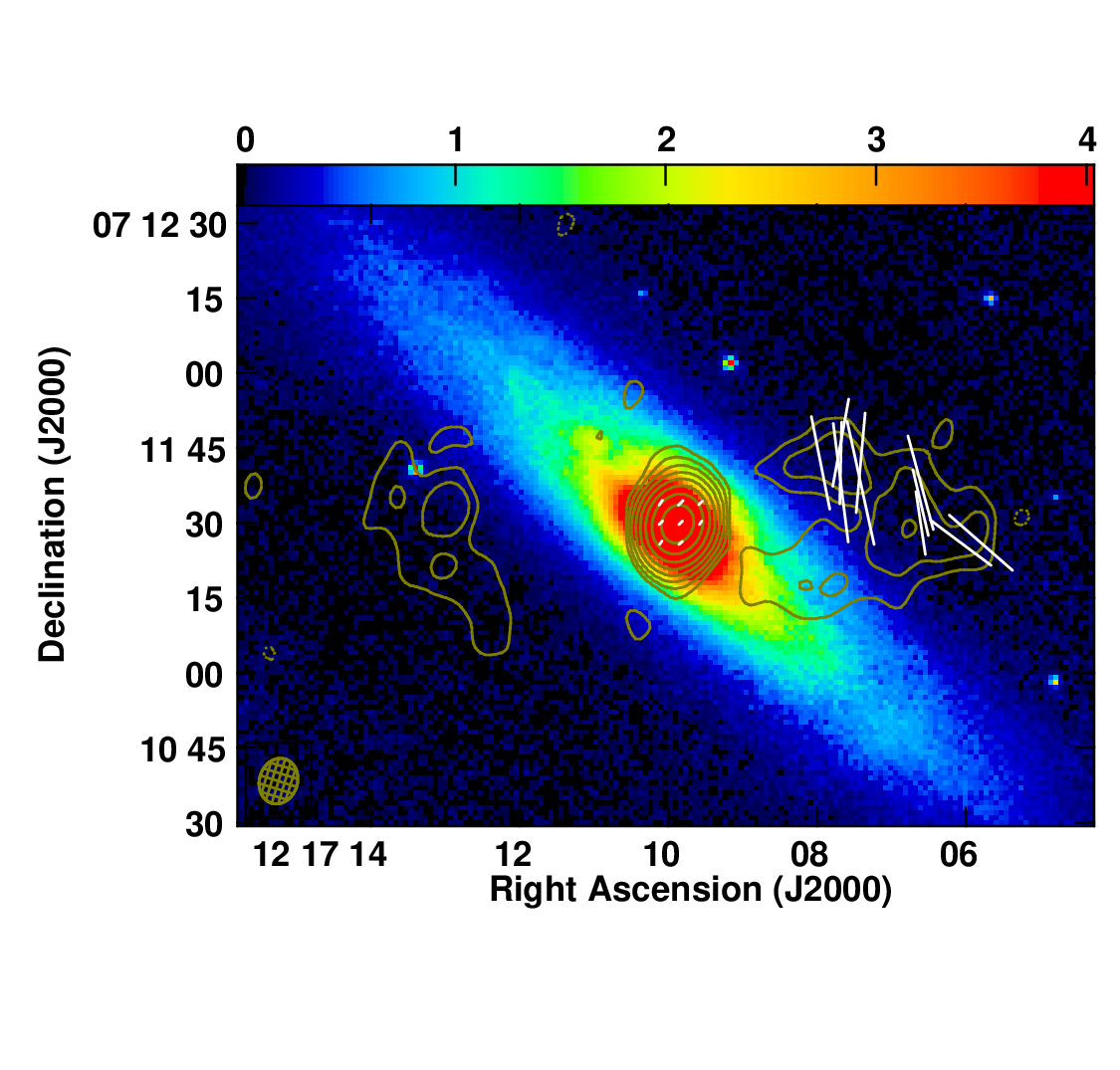}
  \put(75,70){\Large\textbf{VLA 10 GHz}}
  \end{overpic}
  \hfill
  \includegraphics[width=6.3cm, trim= 10 0 10 40,clip]{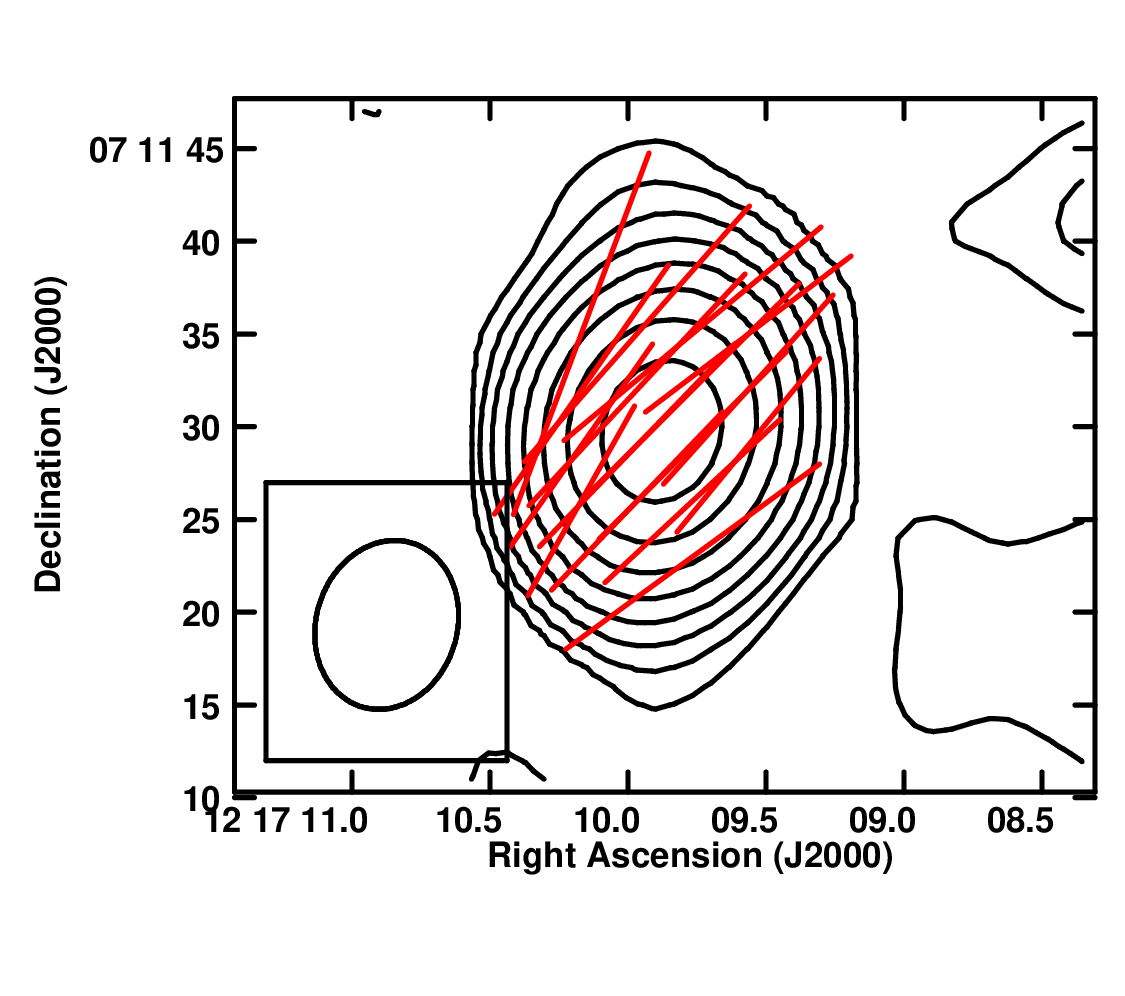} 
  \vspace{0.7cm}
  \newline
  \begin{overpic}[width=11cm, trim= 0 40 10 0,clip]{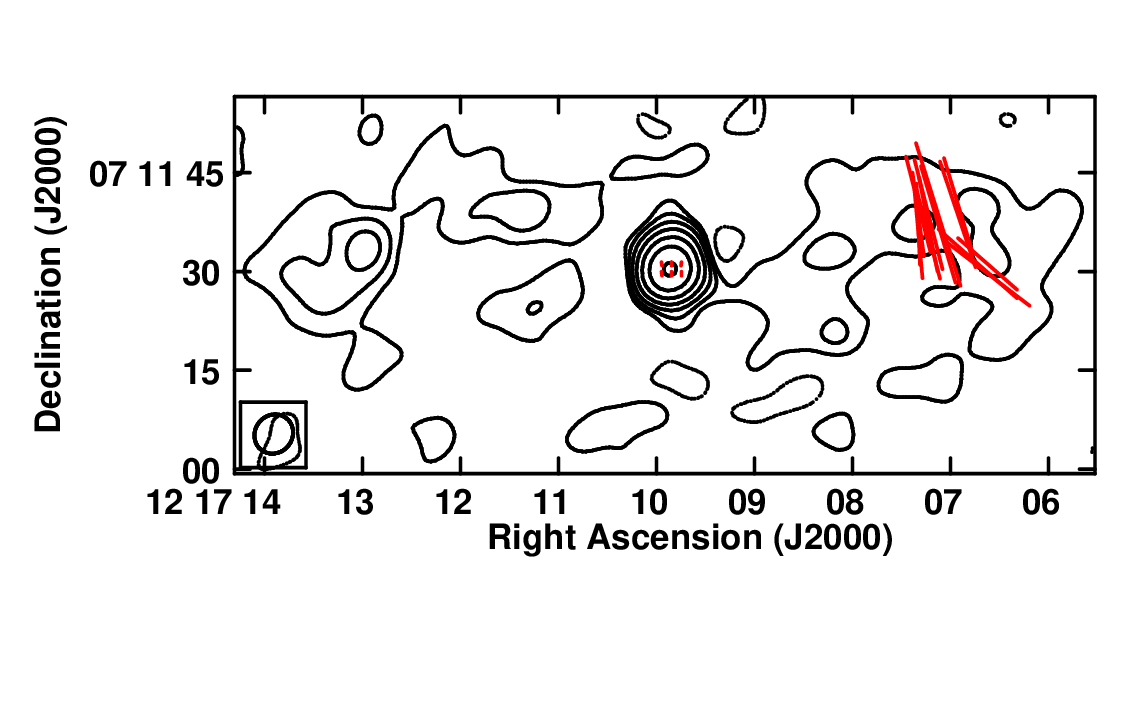}
  \put(70,56){\Large\textbf{uGMRT 1.3 GHz}}
  \end{overpic}
  \hfill
  \includegraphics[width=6.3cm, trim= 15 50 10 40,clip]{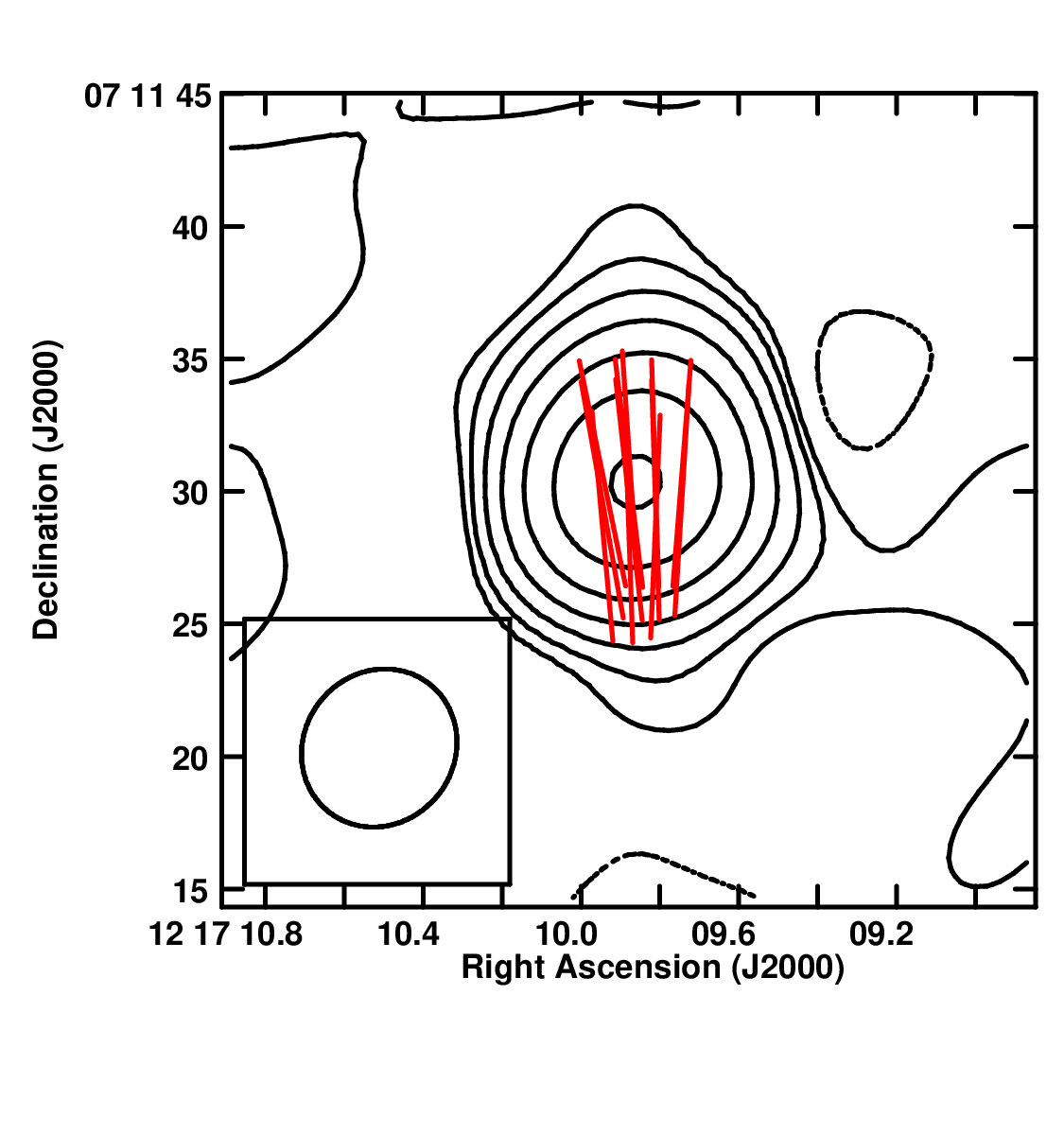}
  \vspace{0.7cm}
  \newline
  \begin{overpic}[width=11cm, trim= 0 40 10 0,clip]{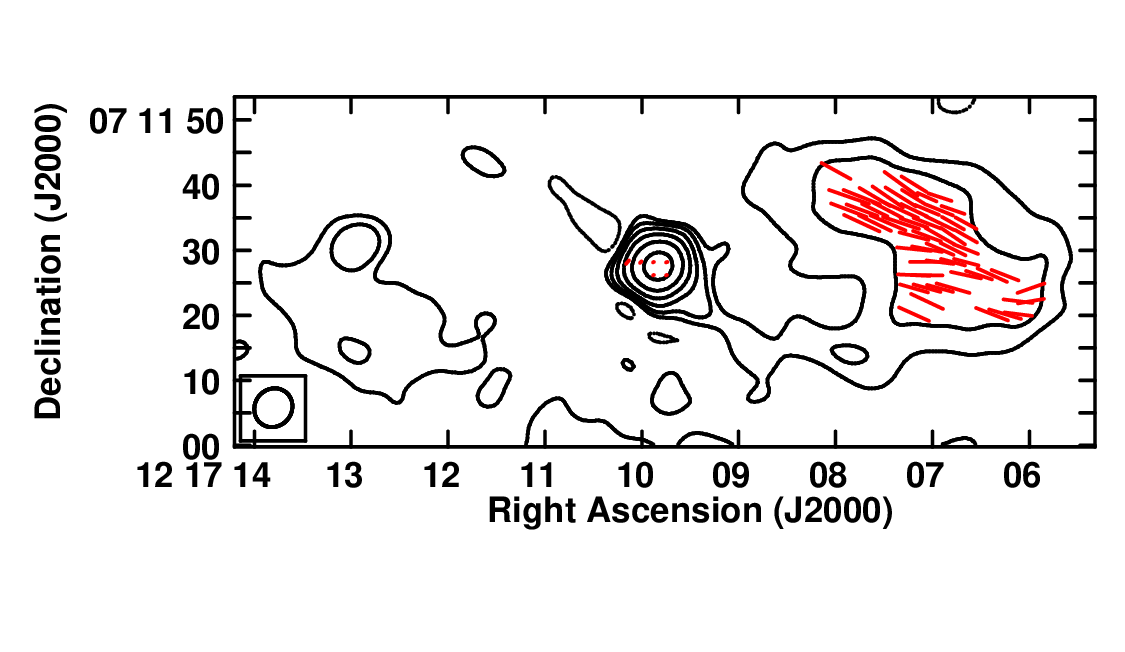}
  \put(70,52){\Large\textbf{uGMRT 0.66 GHz}}
  \end{overpic}
  \hfill
  \includegraphics[width=6.3cm, trim= 15 50 10 40,clip]{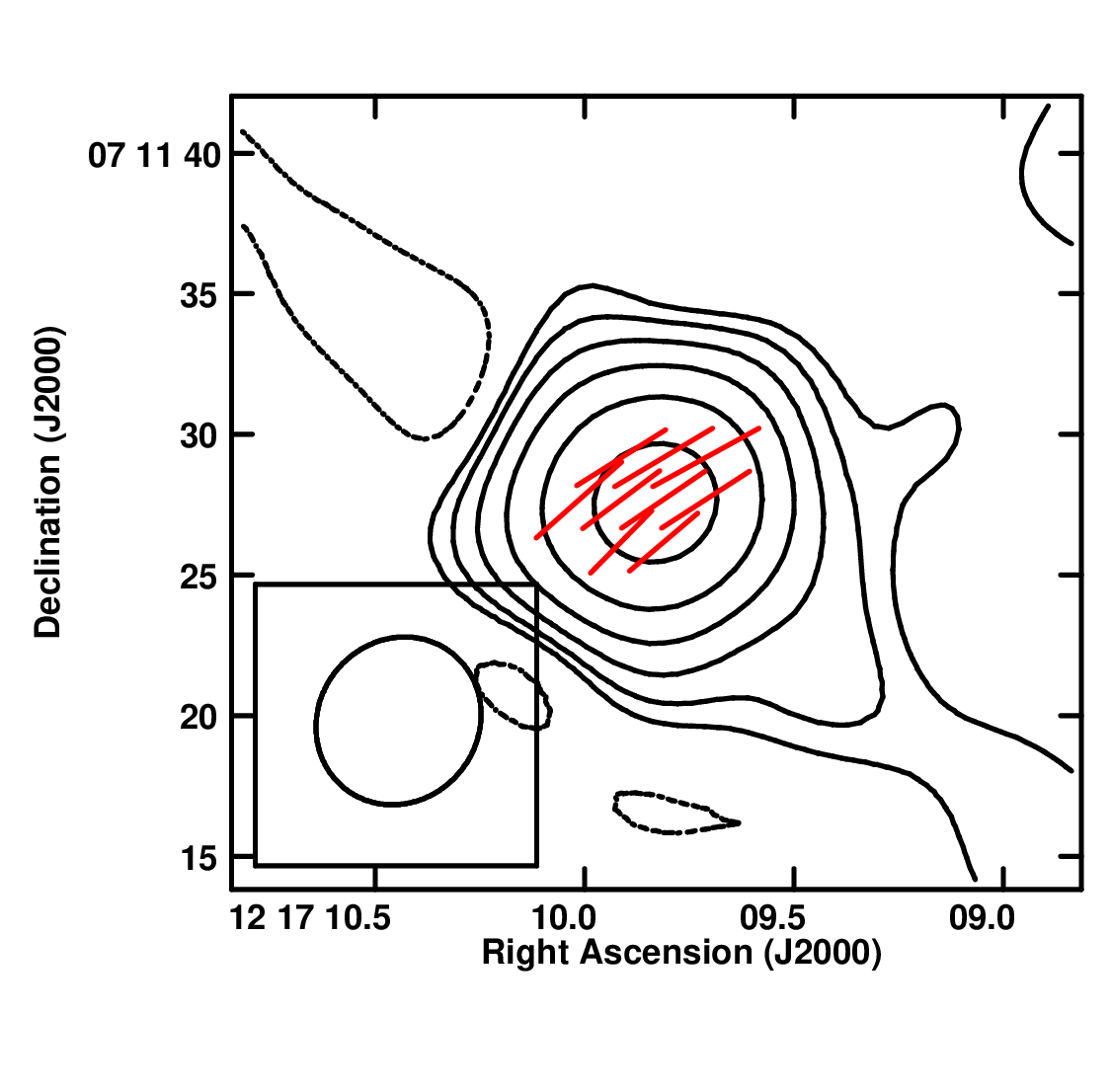}

\caption{\small From top to bottom: VLA 10 GHz contours in olive green overlaid on the Pan-STARRS g-band optical image of NGC~4235 with white EVPA vectors, uGMRT 660 MHz black contours with red EVPA vectors, uGMRT 1.3 GHz black contours with red EVPA vectors. Zoomed-in images of the cores at the respective frequencies are shown on the right panel. {All the EVPA vectors shown have been scaled by fractional polarization with lengths of 10$^{\prime\prime}$ equivalent to 25\% in the left panels and 1.7\% at all frequencies in the right panels.} The synthesized beam is shown as an ellipse in the lower-left corner of each panel (see Table~\ref{tab:radioprop}). Radio contours are at $3\sigma \times (\pm1, 2, 4, 8, 16, 32, 64, 128)$, where $\sigma = 10, 50,$ and $60~\mu$Jy~beam$^{-1}$ at 10 GHz, 1.3 GHz, and 660 MHz, respectively. North is up and east is to the left.}

\label{fig:EVPA_PPOL}
\end{figure*}

In a circular basis, the polarization calibration includes cross-hand delay calibration, leakage (or D-term) calibration, and cross-hand phase calibration solved using {\tt CASA} tasks {\tt gaincal} and {\tt polcal}. We have used the polarized calibrator 3C 286 for polarization angle, delay and leakage calibration. For the circular basis, we have also verified the leakage solutions using an unpolarized calibrator, 3C 147. {In the case of the linear basis, after running the basic calibration, we perform a gain calibration using a model of an unpolarized calibrator of flux density unity. The assumption of the polarized source being unpolarized helps in absorbing into the gains the linear polarization rotation with the parallactic angle, which changes with time. We derive the calibrator's polarization information (Stokes Q and U) from the gains using the function {\tt polfromgain}. Then we calibrate for the cross-hand delays to minimise the phase correction errors. This is followed by the application of the bandpass, initial gain and cross-hand delay calibrations. The XY-phase offset, and the Q, U values from the cross-hands for the calibrator are then estimated using `{\tt Xfparang+QU} solve' in the task {\tt polcal}. The solution is then applied, and the gain calibration is repeated, now with true Stokes values by using the full polarization model derived in the previous step. The instrumental leakages are then solved using the task {\tt polcal} with {\tt poltype = `Dflls'}. The cross-hand gain amplitudes are then normalised. Finally, all the calibration tables are applied, initially to the polarized calibrator and later to the phase calibrator and the target. The calibrated target {\tt uvdata} are then fast Fourier transformed into images using the task {\tt tclean}. 

This is followed with four rounds of phase-only and three rounds of amplitude+phase self-calibration.} The final outputs from the data reduction procedure include Stokes I, Q, U, and V images. The linear polarization intensity is computed as $I_p = \sqrt{Q^2 + U^2}$, the nominal polarization position angle as $\phi = \frac{1}{2}\tan^{-1}(U/Q)$, and the fractional polarization $f_p$ as $I_p/I$.  {These are carried out by using the {\tt COMB} task in {\tt Astronomical Image Processing System} \citep[AIPS;][]{Greisen1990} with {\tt opcode = `POLC'}, {\tt = `POLA'}, and {\tt = `DIV'}, respectively}. { Pixels with errors above $3\sigma$ in $I_p$, 20\% error in $f_p$, and 10$\degree$ error in $\phi_{\rm{obs}}$, are blanked in all the images. The uncertainties are obtained from the error images generated by {\tt AIPS} by propagating errors from the Stokes I, Q, U images.} The final polarization images at 10 GHz, 1.3 GHz and 660 MHz are presented in Figure~ \ref{fig:EVPA_PPOL} in the top, middle, and bottom panels, respectively, with the EVPA vectors having lengths proportional to $I_p$, and angle $\phi_{\rm{obs}}$. All the values for different regions at different frequencies have been tabulated in Table \ref{tab:radioprop}.

An `XY' swap with respect to the IAU/IEEE convention has been pointed out for the uGMRT feeds for Band-4 \citep{Das2020}. {The EVPAs in Figure \ref{fig:EVPA_PPOL}, middle panel, and Table \ref{tab:radioprop} have been presented after rotating the originally obtained values in Band-5 by 90$\degree$. We find that this rotation provides us consistent results with the expected values of $\phi$ for well studied sources like 3C~286 and 3C~273. The 90$\degree$ rotation is consistent with a swap and negation in the linear feeds ({\color{blue}EVLA\_Memo 219\footnote{\href{https://library.nrao.edu/public/memos/evla/EVLAM_219.pdf}{https://library.nrao.edu/public/memos/evla/EVLAM\_219.pdf}}; Technical report by Ghosh et al., in preparation).} Note that the polarization angle calibration in Band-4 with a known angle calibrator like 3C~286 and 3C~138 takes care of any swap in the orthogonally polarized signals \citep[e.g.,][]{Ghosh2023}.}

{We have made an RM image (see Figure \ref{fig:RMimage}) using the VLA 10 GHz Stokes Q, U data by splitting the 4~GHz wide data into three sub-bands (8.68, 9.87, and 11.2 GHz). For a consistent multi-frequency analysis, Stokes $I$, $Q$, and $U$ images at all bands were produced with identical image size, cell size, and a common restoring beam. To achieve a matched angular resolution of $10.3''\times9.1''$, PA=$-8.1\degree$, we adjusted the {\tt robust} parameter for Briggs weighting and applied appropriate {\tt uvrange} limits in the {\tt tclean} imaging task. QU cubes at the three frequencies were used to obtain the RM image using the task {\tt rmfit}. 

We aimed to constrain the RM to the lower uGMRT frequencies at the `core'\footnote{Defined as the compact central radio feature having a size roughly equal to or less than the FWHM of the synthesised beam and comprising primarily the emission from the unresolved jet.}, which is well-fitted with a single Gaussian component\footnote{using the task {\tt imfit} in {\tt CASA}} having a size of $\sim9''$ or $\sim2$ kpc in the 10 GHz VLA image and $\sim6''$ or $\sim1$ kpc in the 660 MHz GMRT image. The EVPAs are therefore obtained at five frequencies spanning 0.66$-$11.2 GHz (see Table \ref{tab:5freqEVPA}). Three of the five frequencies are closely spaced in $\lambda^2$, while the remaining two are widely separated from each other. EVPA is only measured modulo $\pi$ (i.e. 180$^\circ$), each observed value $\phi_{\rm{obs},i}$ is degenerate with $\phi_{\rm{obs},i}+n_i\pi$ for any integer $n_i$ \citep[e.g.,][]{Algaba2011, Liu2017}. This is known as `n$\pi$ ambiguity' in RM fitting. Hence the obtained RM by a linear fit to the raw EVPA values is only valid if $n_i=0$. The task {\tt rmfit} uses at least three well-separated frequency bands to resolve the potential n$\pi$ ambiguities in the polarization angle by $\chi^2$ minimisation, and to obtain a unique solution of RM for each pixel \citep{Leahy1986}. 
For the manually performed linear-fitting, we have included the uGMRT frequencies (i.e., the longer wavelength regime) to put better constraints on the RM. We have assumed that the EVPA has a linear dependence on wavelength-squared without any curvature. We ran our {\tt PYTHON} script to search for all possible $n_i$ values and performed a global $\chi^2$ minimisation to obtain the RM value. The search radius was $-2000<RM<2000$ and $n_i\leq50$ 
\citep[these are constrained by typical RM values observed in the kpc-scale lobes of radio-loud AGN; e.g.,][]{Kharb2008}. The $n_i$ values were computed relative to the $8-12$ GHz data. For every trial combination, an inverse-variance-weighted linear least-squares fit of $\phi$ vs $\lambda^2$ was performed and $\chi^2$ computed, as presented in Table \ref{tab:RMsoln}.}


\section{Results \& Discussion}
NGC~4235 exhibits extended radio emission with an overall S-shaped/bubble morphology \citep[as often observed in Seyfert galaxies;][]{Kharb2006,Kharb2010}. Considering all of the extended S-shaped emission is related to the AGN, we detect a total flux density of $17.1\pm1.7$ mJy at 1.3 GHz, $27.7\pm2.8$ mJy at 663 MHz. The extraplanar radio lobes appear asymmetric, with the brighter western lobe likely located on the near side of the galactic disk, while the fainter eastern lobe, which shows evidence of a terminal hotspot, is situated behind the disk and subject to increased  depolarization. The radio `core' at 10 GHz shows an $\alpha$ of $+0.11\pm0.02$ \citep{Ghosh2025}, consistent with a flat to inverted spectrum, suggesting that the nuclear emission is optically thick due to synchrotron self-absorption. In such optically thick regions, the projected magnetic field is expected to be oriented parallel to the observed EVPAs \citep[see][]{Contopoulos2015}. {The EVPAs at 10 GHz indicate that the B-field in the inner optically thick core} is predominantly perpendicular to the outflow axis, consistent with a toroidal or helical B-field configuration. Such a B-field geometry is expected in wide-angle disk winds or magnetically driven outflows launched from the accretion disk, which can collimate a weak jet from parsec to kpc scales \citep[see][]{Mehdipour2019}.

\begin{figure}
  \centering
  \includegraphics[width=7.5cm, trim= 50 20 30 0]{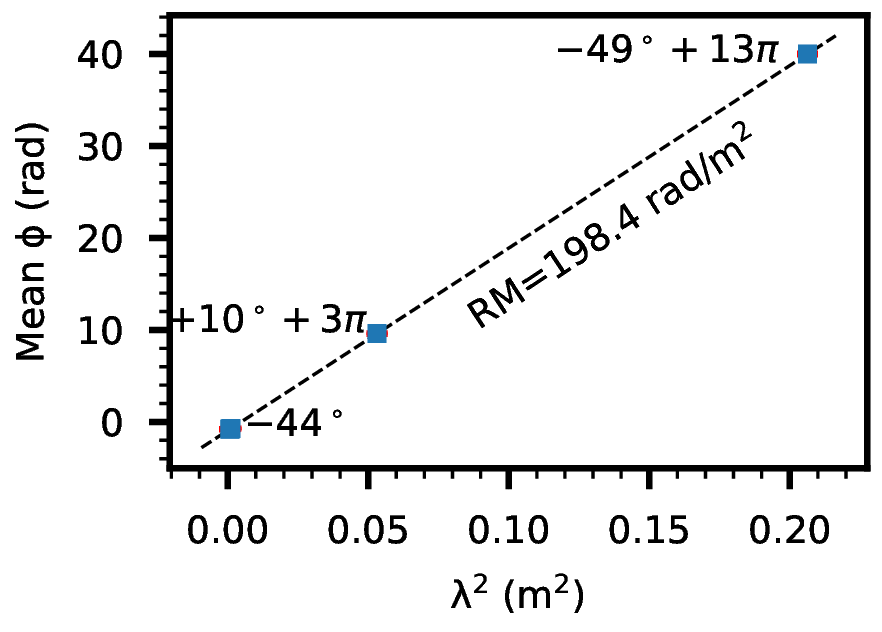}
  \caption{\small {Mean polarization angle or EVPAs plotted as a function of wavelength squared. The EVPAs in the y-axis are n$\pi$ de-rotated values for the lowest $\chi^2$ fit. The errors in the EVPAs are shown in red, which are smaller than the size of the markers. The black dashed line shows the best-fit with a slope of $198.4\pm0.7$ and an intercept of $-0.95\pm0.05$.}} 
  \label{fig:RMnpi}
\end{figure}

{
As described in Section \ref{sec:obs}, we searched through different n-values to fit the EVPAs ($\phi+$n$\pi$) over a range of frequencies for the `core'. The global minimum of $\chi^2$ was obtained for branch offsets (relative to the $\sim8-12$ GHz data) of n$=0,0,0,+3,+13$ at 8.68, 9.87, 11.2, 1.3, and 0.66 GHz, respectively. The corresponding best-fit parameters are $RM=198.4\pm0.7$ rad m$^{-2}$, and intrinsic polarization angle $\phi_{\rm{int}}=-54.5^\circ\pm2.9^\circ$, with $\chi^2=0.0086$ for N-2=3 degrees of freedom, and reduced $\chi^2$ ($\chi_\nu^2$) = 0.0029. The $\phi$ vs. $\lambda^2$ plot for the best-fit has been shown in Figure \ref{fig:RMnpi}. 
All residuals for the best-fit solution are $<0.4^\circ$, well within the quoted measurement uncertainties ($5-8^\circ$), supporting our conclusion that the obtained RM solution is correct and unique.}

\begin{table*}
\caption{\small Polarization properties of NGC 4235 at different frequencies.} \label{tab:radioprop}
{\scriptsize
\begin{tabular}{cccc|cccc|ccc} \hline
Telescope&Frequency&Beam size&rms noise (I, Q, U)&\multicolumn{4}{|c|}{Core}&\multicolumn{3}{c}{Western lobe}\\
& (GHz) &($'','',^\circ$)&($\mu$Jy beam$^{-1}$) &$I (mJy)$&$I_p (mJy)$&$f_p (\%)$&$\phi (^{\circ})$&$I (mJy)$&$I_p (mJy)$&$f_p (\%)$\\ \hline
VLA&10&$9.2\times7.6, -17.2$&$10, 9, 9$&$6.3\pm0.2$&$0.13\pm0.02$&$2.7\pm0.5$&$-43\pm5$&$1.1\pm0.1$&$0.16\pm0.04$&$54\pm16$ \\
uGMRT&1.3&$6.1\times5.7, -38.8$&$50, 35, 35$&$7.8\pm0.1$&$0.13\pm0.04$&$1.4\pm0.4$&$+10\pm7$&$4.7\pm0.7$&$0.33\pm0.09$&$44\pm15$\\
uGMRT&0.66&$6.1\times5.7, -38.8$&$60, 15, 13$&$9.2\pm0.2$&$0.022\pm0.006$&$0.9\pm0.3$&$-49\pm8$&$13\pm2$&$1.3\pm0.3$&$16\pm15$\\
\hline
\end{tabular}}
{\small Note: Column 1, 2, and 3: Telescope name, observing frequency. and synthesized beam sizes. Column 4: rms noise for Stokes I, Q, U images. Column 5, 6, 7, and 8: Flux density, polarized flux density, mean fractional polarization, and mean polarization angle in degrees. Column 9, 10, and 11: Flux density, polarized flux density, and mean polarization fraction for the Western lobe; the polarization angles have not been noted as the lobe shows multiple polarized regions with different angles.}
\end{table*}

\begin{table}
{\caption{Observed EVPAs at different frequencies.}
\label{tab:5freqEVPA}
\begin{tabular}{ccc}
\hline
Frequency (GHz) & $\lambda^2$ (m$^2$) & $\phi_{\rm{obs}}$ $(^\circ)$\\
\hline
0.66  & $2.0633\times10^{-1}$ & $-49.0\pm8.0$\\
1.30  & $5.3181\times10^{-2}$ & $+10.0\pm7.0$\\
8.68  & $1.1929\times10^{-3}$ & $-41.3\pm5.0$\\
9.87  & $9.2259\times10^{-4}$ & $-44.0\pm5.0$\\
11.20 & $7.1648\times10^{-4}$ & $-46.0\pm6.0$\\
\hline
\end{tabular}
{\small Note: Column 1: Frequency of observations in GHz. Column 2: Corresponding wavelength-squared in m$^2$, calculated as the speed of light divided by the observing frequencies (in column 1) whole squared. Column 3: Observed EVPA values 
in degrees.}}
\end{table}

\begin{table}
{\caption{Fitted RM-solutions from the $n\pi$-ambiguity search, ranked in increasing order of $\chi^2$. The offsets $n_{1.3}$ and $n_{0.66}$ provide the number of $\pi$-turns applied to the 1.3 and 0.66~GHz points relative to the 8.68/9.87/11.2 GHz data, which is fixed at $n_i=0$. }
\label{tab:RMsoln}
\setlength{\tabcolsep}{3pt}
\begin{tabular}{ccccc}
\hline
RM (rad\,m$^{-2}$) & $\phi_{\rm{int}}$ ($^{\circ}$) & $n_{1.3\,\rm GHz}$, $n_{0.66\,\rm GHz}$ & $\chi^2$ & $\chi^2_\nu$\\
\hline
$198.411$  & $-54.54$ & $+3$, $+13$ & $0.009$ & $0.003$ \\
$137.260$  & $-50.78$ & $+2$, $+9$  & $0.173$ & $0.058$ \\
$259.562$  & $-58.30$ & $+4$, $+17$ & $0.259$ & $0.087$ \\
$76.109$   & $-47.02$ & $+1$, $+5$  & $0.753$ & $0.252$ \\
$320.713$  & $-62.06$ & $+5$, $+21$ & $0.926$ & $0.309$ \\
$14.958$   & $-43.26$ & $0$, $+1$  & $1.749$ & $0.583$ \\
\hline
\end{tabular}
{\small Note: Column 1: Fitted RM values in rad~m$^{-2}$. Column 2: $\phi_{\rm{int}}$ is the RM-corrected EVPA. Column 3: The number of n$\pi$ turns for the EVPAs at 1.3 GHz and 660 MHz, keeping the EVPA at $\sim$10 GHz constant. Column 4: $\chi^2$ is the measure of goodness-of-fit statistic, obtained as the inverse-variance-weighted sum of
squared residuals between the n$\pi$ de-rotated position angles and the best-fit EVPAs, $\chi^2 \;=\; \sum_{i=1}^{N} \left( \frac{\phi_i - \phi_{0,i}}{\sigma_i} \right)^{2}$, where $\phi_i = \phi_{{\rm obs},i} + n_i\pi$, $\phi_{0,i} = \phi_{\rm{int}} + \mathrm{RM}\,\lambda_i^2$, and $\sigma_i$ is the error on each EVPA measurement. Column 5: Reduced $\chi^2$ measured as $\chi^2$ divided by the degrees of freedom which is 3 here (N-2, where N = 5 for the number of data points).}}
\end{table}

\begin{table}

{\caption{Best-fit RM model for the `core'.}
\label{tab:evpa_fit}
\setlength{\tabcolsep}{3pt}
\begin{tabular}{ccccc}
\hline
Frequency (GHz) & $n_i$ & $\phi_{\rm corr}$ ($^\circ$) & $\phi_{\rm model}$ ($^\circ$) & Residual ($^\circ$) \\
\hline
0.66  & $+13$ & $2291.00$ & $2290.99$ & $+0.01$ \\
1.30  & $+3$  & $550.00$ & $550.02$ & $-0.02$ \\
8.68  & 0   & $-41.30$ & $-40.98$ & $-0.32$ \\
9.87  & 0   & $-44.00$ & $-44.05$ & $+0.05$ \\
11.20 & 0   & $-46.00$ & $-46.40$ & $+0.40$ \\
\hline
\end{tabular}
{\small Note: Column 1: Observing Frequencies in GHz. Column 2: n$\pi$ turns added to the $\phi_{\rm obs}$ for the different frequencies. Column 3: Corrected $\phi_{\rm obs}$ after n$\pi$ rotation. Column 4: Model $\phi_{\rm obs}$ from the best fit solution of RM is $198.4\pm0.7$ rad m$^{-2}$. Column 5: Residuals estimated as the difference between $\phi_{\rm corr}$ and $\phi_{\rm model}$.}}
\end{table}


Early polarimetric studies of extragalactic radio sources established that kpc-scale jets and lobes typically exhibit $f_p$ of $\sim10-50$\% and RMs ranging from a few tens up to several hundred rad~m$^{-2}$, with values increasing toward host galaxies and dense environments. This was considered to be consistent with Faraday rotation arising in external magneto-ionic media rather than within the synchrotron-emitting plasma itself \citep{Saikia1988}. Therefore, the modest RM value for NGC 4235 suggests that {there is an absence of a strong depolarization in the `core', indicating that the Faraday-rotating medium is thin and uniform.}

NGC~4235 shows ordered B-field structures aligned with the kpc-scale radio outflow and high fractional polarization in the Western lobe, indicating coherent magnetic fields similar to those observed in classical FRI radio galaxies, despite the source being a radio-quiet AGN. The eastern lobe is depolarized, possibly due to the Laing-Garrington effect \citep[e.g.,][]{Laing2008}. The consistency between the VLA-scale ordered magnetic fields and the uGMRT-scale RM values indicates that the kpc-scale radio emission in NGC~4235 is Faraday-rotated by the magnetized plasma in the host galaxy's interstellar medium.

\begin{figure}
  \centering
  \includegraphics[width=0.5\textwidth]{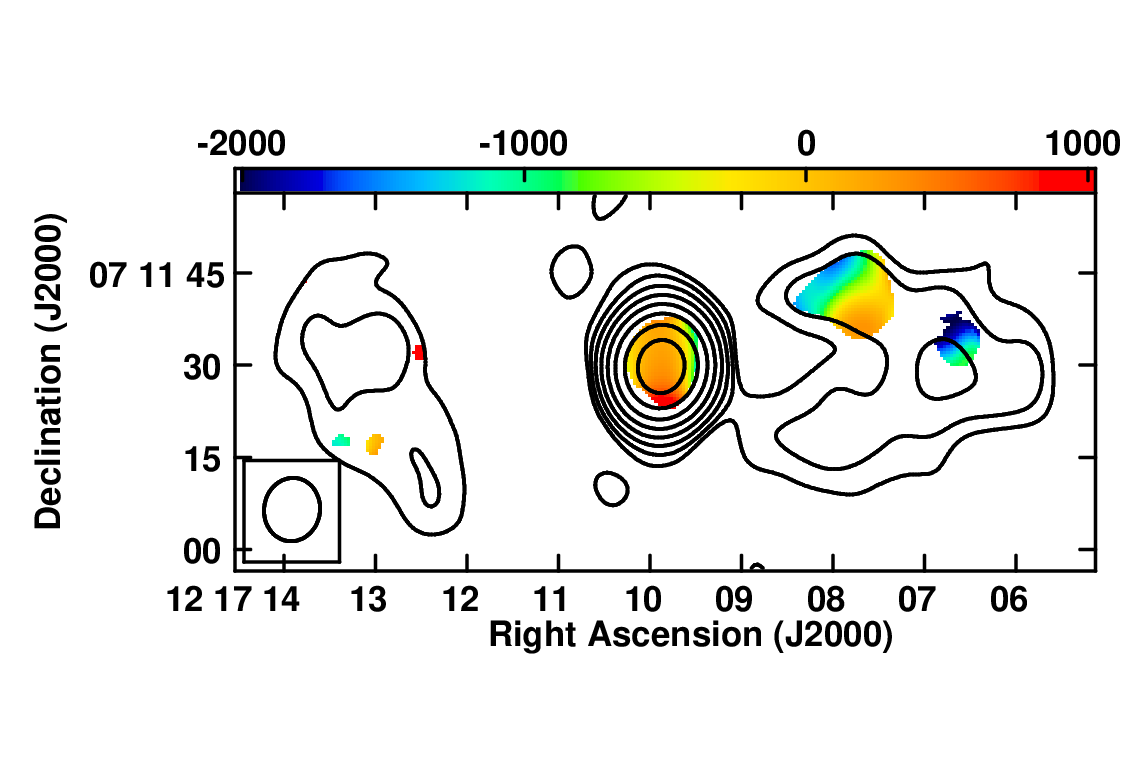}
  \caption{\small VLA in-band ($8-12$ GHz) RM image of NGC~4235 created using the task {\tt rmfit} in CASA. The RM value shown in colour ranges from -2000 to +1000 rad m$^{-2}$.}
  \label{fig:RMimage}
\end{figure}

We find that the RM obtained from higher frequencies (8$-$12~GHz) using only the VLA (obtained by splitting the 4 GHz bandwidth data into three sub-bands), {as shown in Figure \ref{fig:RMimage}}, yields larger values, ranging from $-2000$ to $+1000\ \mathrm{rad\ m^{-2}}$, with a `core' RM of $\sim200\ \mathrm{rad\ m^{-2}}$. {When the RM is constrained using the VLA 10 GHz, together with the GMRT 1.3 GHz and 660 MHz polarization measurements, the inferred n$\pi$-corrected RM remains similar. These results demonstrate that a single value of RM is responsible for the observed EVPAs over a wide range of frequencies in NGC~4235.} 

These results highlight the reliability of uGMRT polarimetric measurements, which provide more constraints on the polarimetric modelling and better information about AGN jets and their environments. The Galactic RM along this line of sight is found to be $5.5\pm1.5$ rad m$^{-2}$ \citep{Xu2014}. Subtracting this from the observed RM at the `core', the resultant `core' RM becomes {$192.9\pm1.7$ rad m$^{-2}$}. The RM de-rotated polarization angle at the `core' is found to be $\sim-54.54^\circ$ or $\sim126^\circ$ using the CASA task {\tt rmfit}.

{For an estimate of the magnetic field strength, we adopt the standard equipartition assumption commonly used in radio-emitting AGN \citep{OdeaOwen1987}. We estimate the B-field to be $\sim5-8~\mu$G in the core using both the VLA 10 GHz and uGMRT 660 MHz data. Assuming that the characteristic line-of-sight field strength is the same as the equipartition B-field \citep[see][]{Feain2009, Kharb2009},} substituting these values into Eqn.~\ref{eq:RM}, and assuming a uniform magneto-ionic medium along our line of sight, we infer an average electron density of {$n_e=(2.4-2.7)\times10^{-2}$ cm$^{-3}$}. 
This is obtained considering the previously derived resultant `core' RM and the length of the Faraday-rotating medium to be equivalent to the size of the `core', with the range involving different values of core-sizes and B-field strengths for both the VLA and uGMRT data.  {The emission in the western lobe contains several polarized components with different EVPAs. Therefore, we could not derive a single integrated RM for the lobe. The RM image in Figure \ref{fig:RMimage} shows RM changing from -1000 to +500 rad m$^{-2}$ across the western lobe with a sharp transition from negative to positive values from the north to the south. This is indicative of the electric vectors undergoing a rotation along the direction of the bending of the lobe.}

{For the `core', the excellent linearity of the $\phi$ vs $\lambda^2$ plot across the broad frequency range provides evidence for Faraday rotation by an external screen. The $\phi-\lambda^2$ relation shows no significant curvature, with a reduced $\chi^2 \sim 0.003$. In case of an internal Faraday rotation within the synchrotron-emitting plasma, it could introduce wavelength-dependent depolarisation and deviation from this simple linear behaviour when the Faraday depth across the emitting region is sufficiently large \citep{Burn1966,Sokoloff1998}. Large EVPA variations between physically distinct components
together with smoother variations within individual components, are consistent with the radio emission 
traversing different paths through magneto-ionised gas that is external to the radio-emitting source.} The external Faraday rotation and depolarization conditions have been noted by \citet{Burn1966} as
\begin{equation}
    p(\lambda^2)=p_i\,\exp{\{-2\,(812\,n_eB_\parallel)^2\,d\,R\,\lambda^4\}}
\label{eq:externaldepolarization}
\end{equation}
where $p_i$, $R$, $d$ are the intrinsic fractional polarization, depth of the Faraday-rotating cloud along the line of sight in kpc, and the perturbation-scale in kpc causing the rotation. $B_\parallel$ and $\lambda$ are in $\mu$G and m, respectively. Using this relation, the best-fitting solution for the `core' from our multi-frequency polarization data, with $R$ comparable to the size of the observed radio core, yields an intrinsic polarization fraction value of {$2.29\pm0.37$\% and the perturbation scale length of $0.32^{+0.20}_{-0.13}$} pc. {As the `core' corresponds to the unresolved base of the jet  \citep[e.g.,][]{Blandford1979}, the most plausible site for such a screen is a compact, magnetized sheath surrounding the jet itself \citep[also see][]{Zavala2004}.} It is interesting to note that the perturbation scale-length of {$\sim$0.3 pc in a region of $\sim$1.3 kpc} can correspond to a thin magnetized sheath, or perhaps an AGN-driven wind surrounding the radio outflow \citep[e.g.,][]{Ghosh2026}.   

The residuals after fitting are small and show no systematic trend with wavelength, and they are tabulated in Table \ref{tab:evpa_fit}. Given the limited number of data points and the assumption of a single-component external depolarization model, this may reflect a combination of unaccounted systematic uncertainties and intrinsic complexity in the magneto-ionic medium along the line of sight.

\section{Summary}
We have obtained the first-ever rotation measure in the core and lobes of a Seyfert galaxy, viz., NGC~4235, using the uGMRT at 660~MHz (Band-4) and 1.3~GHz (Band-5), along with the VLA 10 GHz polarization data. An RM of {$192.9\pm1.7$~rad~m$^{-2}$} suggests a low electron density ({$\approx3\times10^{-2}$ cm$^{-3}$}) and small {perturbation length scale ($\sim0.3$ parsec) within an assumed magnetized region of $1.1-1.5$ kpc}, for the Faraday-rotating medium. These properties are consistent with a diffuse magnetized sheath or AGN-driven wind surrounding the radio jet-like outflow. The relatively small RM further indicates that significant internal Faraday depolarization within the core is unlikely, suggesting that the observed Faraday rotation predominantly originates in an external, low-density magneto-ionized medium.

\section{Acknowledgements}
SG, PK acknowledge the support of the Department of Atomic Energy, Government of India, under the project 12-R\&D-TFR-5.02-0700. JB acknowledges financial support from the Inter University Institute for Data Intensive Astronomy (IDIA). IDIA is a partnership of the University of Cape Town, the University of Pretoria and the University of the Western Cape. IDIA is registered on the Research Organization Registry with ROR ID 01edhwb26, and on OpenFunder Registry with funder ID 100031500. This research has made use of data from the National Radio Astronomy Observatory (NRAO) facility. The NRAO is a facility of the National Science Foundation operated under cooperative agreement by Associated Universities, Inc. We thank the staff of the GMRT who made these observations possible. GMRT is run by the National Centre for Radio Astrophysics of the Tata Institute of Fundamental Research.

\bibliographystyle{apj}
\bibliography{ms}

\end{document}